\documentclass[12pt]{article}
\usepackage{scicite}
\let\citep=\cite

\usepackage{graphicx}
\usepackage{authblk}
\usepackage{lineno}
\usepackage{paralist}
\usepackage{amsmath}

\newenvironment{sciabstract} 
{\bfseries}
{}

\title{Non-equilibrium evolution of volatility in origination and
  extinction explains fat-tailed fluctuations in Phanerozoic
  biodiversity \\
  \vspace{1em} \large {\it One sentence summary:} Phanerozoic marine
  invertebrate richness fluctuates out of equilibrium due to pulsed
  adaptive evolution.}

\author[1, {*}]{Andrew J. Rominger}
\author[1, 2, 3]{Miguel A. Fuentes}
\author[1, 4, 5, 6, 7, 8]{Pablo A. Marquet}

\affil[1]{\normalsize{Santa Fe Institute, 1399 Hyde Park Road, Santa Fe, New
Mexico 87501, US}}
\affil[2]{\normalsize{Instituto de Investigaciones Filos\'oficas, SADAF, CONICET,
Bulnes 642, 1428 Buenos Aires, Argentina}}
\affil[3]{\normalsize{Facultad de Ingenier\'ia y Tecnolog\'ia, Universidad San
Sebasti\'an, Lota 2465, Santiago 7510157, Chile}}
\affil[4]{\normalsize{Departamento de Ecolog\'ia, Facultad de Ciencias
Biol\'ogicas, Pontificia Universidad de Chile, Alameda 340, Santiago,
Chile}}
\affil[5]{\normalsize{Instituto de Ecolog\'ia y Biodiversidad (IEB),
    Casilla 653, Santiago, Chile}}
\affil[6]{\normalsize{Laboratorio Internacional de Cambio Global
    (LINCGlobal), and Centro de Cambio Global UC, Pontificia
    Universidad Catolica de Chile, Santiago, Chile.}}
\affil[7]{\normalsize{Instituto de Sistemas Complejos de Vlapara\'iso
    (ISCV), Artiller\'ia 470, Cerro Artiller\'ia, Valpara\'iso,
    Chile}}
\affil[8]{\normalsize{Centro de Ciencias de la Complejidad (C3),
    Universidad Nacional Aut\'onoma de M\'exico.}}
\affil[{*}]{\normalsize{To whom correspondence should be addressed,
    e-mail: rominger@santafe.edu}}

\date{}

\RequirePackage[normalem]{ulem} %DIF PREAMBLE
\RequirePackage{color}\definecolor{RED}{rgb}{1,0,0}\definecolor{BLUE}{rgb}{0,0,1} %DIF PREAMBLE
\providecommand{\DIFdel}[1]{{\protect\color{red}\sout{}}}                      %DIF PREAMBLE
\providecommand{\DIFaddbegin}{} %DIF PREAMBLE
\providecommand{\DIFaddend}{} %DIF PREAMBLE
\providecommand{\DIFdelbegin}{} %DIF PREAMBLE
\providecommand{\DIFdelend}{} %DIF PREAMBLE
\providecommand{\DIFaddbeginFL}{} %DIF PREAMBLE
\providecommand{\DIFaddendFL}{} %DIF PREAMBLE
\providecommand{\DIFdelbeginFL}{} %DIF PREAMBLE
\providecommand{\DIFdelendFL}{} %DIF PREAMBLE
\newcommand{\DIFscaledelfig}{0.5}
\RequirePackage{settobox} %DIF PREAMBLE
\RequirePackage{letltxmacro} %DIF PREAMBLE
\newsavebox{\DIFdelgraphicsbox} %DIF PREAMBLE
\newlength{\DIFdelgraphicswidth} %DIF PREAMBLE
\newlength{\DIFdelgraphicsheight} %DIF PREAMBLE
\LetLtxMacro{\DIFOincludegraphics}{\includegraphics} %DIF PREAMBLE
\newcommand{\DIFaddincludegraphics}[2][]{{\color{blue}\fbox{\DIFOincludegraphics[#1]{#2}}}} %DIF PREAMBLE
\newcommand{\DIFdelincludegraphics}[2][]{% %DIF PREAMBLE
\sbox{\DIFdelgraphicsbox}{\DIFOincludegraphics[#1]{#2}}% %DIF PREAMBLE
\settoboxwidth{\DIFdelgraphicswidth}{\DIFdelgraphicsbox} %DIF PREAMBLE
\settoboxtotalheight{\DIFdelgraphicsheight}{\DIFdelgraphicsbox} %DIF PREAMBLE
\scalebox{\DIFscaledelfig}{% %DIF PREAMBLE
\parbox[b]{\DIFdelgraphicswidth}{\usebox{\DIFdelgraphicsbox}\\[-\baselineskip] \rule{\DIFdelgraphicswidth}{0em}}\llap{\resizebox{\DIFdelgraphicswidth}{\DIFdelgraphicsheight}{% %DIF PREAMBLE
\setlength{\unitlength}{\DIFdelgraphicswidth}% %DIF PREAMBLE
\begin{picture}(1,1)% %DIF PREAMBLE
\thicklines\linethickness{2pt} %DIF PREAMBLE
{\color[rgb]{1,0,0}\put(0,0){\framebox(1,1){}}}% %DIF PREAMBLE
{\color[rgb]{1,0,0}\put(0,0){\line( 1,1){1}}}% %DIF PREAMBLE
{\color[rgb]{1,0,0}\put(0,1){\line(1,-1){1}}}% %DIF PREAMBLE
\end{picture}% %DIF PREAMBLE
}\hspace*{3pt}}} %DIF PREAMBLE
} %DIF PREAMBLE
\LetLtxMacro{\DIFOaddbegin}{\DIFaddbegin} %DIF PREAMBLE
\LetLtxMacro{\DIFOaddend}{\DIFaddend} %DIF PREAMBLE
\LetLtxMacro{\DIFOdelbegin}{\DIFdelbegin} %DIF PREAMBLE
\LetLtxMacro{\DIFOdelend}{\DIFdelend} %DIF PREAMBLE
\DeclareRobustCommand{\DIFaddbegin}{\DIFOaddbegin \let\includegraphics\DIFaddincludegraphics} %DIF PREAMBLE
\DeclareRobustCommand{\DIFaddend}{\DIFOaddend \let\includegraphics\DIFOincludegraphics} %DIF PREAMBLE
\DeclareRobustCommand{\DIFdelbegin}{\DIFOdelbegin \let\includegraphics\DIFdelincludegraphics} %DIF PREAMBLE
\DeclareRobustCommand{\DIFdelend}{\DIFOaddend \let\includegraphics\DIFOincludegraphics} %DIF PREAMBLE
\LetLtxMacro{\DIFOaddbeginFL}{\DIFaddbeginFL} %DIF PREAMBLE
\LetLtxMacro{\DIFOaddendFL}{\DIFaddendFL} %DIF PREAMBLE
\LetLtxMacro{\DIFOdelbeginFL}{\DIFdelbeginFL} %DIF PREAMBLE
\LetLtxMacro{\DIFOdelendFL}{\DIFdelendFL} %DIF PREAMBLE
\DeclareRobustCommand{\DIFaddbeginFL}{\DIFOaddbeginFL \let\includegraphics\DIFaddincludegraphics} %DIF PREAMBLE
\DeclareRobustCommand{\DIFaddendFL}{\DIFOaddendFL \let\includegraphics\DIFOincludegraphics} %DIF PREAMBLE
\DeclareRobustCommand{\DIFdelbeginFL}{\DIFOdelbeginFL \let\includegraphics\DIFdelincludegraphics} %DIF PREAMBLE
\DeclareRobustCommand{\DIFdelendFL}{\DIFOaddendFL \let\includegraphics\DIFOincludegraphics} %DIF PREAMBLE
\begin{document} 

\maketitle 
\clearpage

\begin{sciabstract}
  Fluctuations in biodiversity, both large and small, are pervasive
  through the fossil record, yet we do not understand the processes
  generating them.
  Here we extend theory from non-equilibrium statistical physics to
  describe the previously unaccounted for fat-tailed form of
  fluctuations in marine invertebrate richness through the
  Phanerozoic.
  Using this theory, known as superstatistics, we show that the simple
  fact of heterogeneous rates of origination and extinction between
  clades and conserved rates within clades is sufficient to account
  for this fat-tailed form. We identify orders and the families they
  subsume as the taxonomic level at which clades experience
  inter-clade heterogeneity and within clade homogeneity of
  rates. Following superstatistics we would thus posit that orders and
  families are subsystems in local statistical equilibrium while the
  entire system is not in equilibrium.
  The separation of timescales between background origination and
  extinction within clades compared to the origin of major ecological
  and evolutionary innovations leading to new orders and families
  allows within-clade dynamics to reach equilibrium, while
  between-clade diversification is non-equilibrial.
  This between clade non-equilibrium accounts for the fat-tailed
  nature of the system as a whole.
  The distribution of shifts in diversification dynamics across orders
  and families is consistent with niche conservatism and pulsed
  exploration of adaptive landscapes by higher taxa.
  Compared to other approaches that have used simple birth-death
  processes, simple equilibrial dynamics, or non-linear theories from
  complexity science, superstatistics is superior in its ability to
  account for both small and extreme fluctuations in the richness of
  fossil taxa.
  Its success opens up new research directions to better understand
  the evolutionary processes leading to the stasis of order- and
  family-level occupancy in an adaptive landscape interrupted by
  innovations that lead to novel forms.
\end{sciabstract}

\clearpage

% double-space the manuscript
\baselineskip24pt

\section{Introduction}

Biodiversity has not remained constant nor followed a simple
trajectory through geologic time \citep{raup1982, sepkoski1984,
  gilinsky1994, liow2007, alroy08}.  Instead, it has been marked by
fluctuations in the richness of taxa, both positive in the case of net
origination, or negative in the case of net extinction. Major events,
such as adaptive radiations and mass extinctions have received special
attention \citep{benton1995, Erwin1998}, but fluctuations of all sizes
are ubiquitous \citep{sepkoski1984, alroy08} and follow a fat-tailed
distribution where large events are more probable compared to, e.g. a
Gaussian distribution. Understanding the fat-tailed nature of these
fluctuations continues to elude paleobiologists and biodiversity
theoreticians.

The fat-tailed distribution of fluctuations in taxon richness inspired
earlier researchers to invoke ideas from complex systems with similar
distributions. Such ideas include the hypotheses that biological
systems self-organize to the brink of critical phase-transitions
\citep{bak1993, sole1997} and that environmental perturbations are
highly non-linear \citep{newman1995}. New data and analyses have not,
however, supported these hypotheses at the scale of the entire
Phanerozoic marine invertebrate fauna \citep{kirchner1998, alroy08}.
Other studies have modeled the mean trend in taxon richness as
tracking a potentially evolving equilibrium \citep{sepkoski1984,
  alroy2010, quental2013} and yet ignore the role of stochasticity and
non-equilibrium dynamics in producing observed patterns
\citep{erwin2012, liow2007, jordan2016}. Individual, population, and
local ecosystem scale processes that could produce complex dynamics,
such as escalatory co-evolutionary interactions \citep{vermeij1987},
have not been documented to scale up to global patterns
\citep{madin2006} and indeed should not be expected to do so
\citep{vermeij2008}.  Thus, we still lack a theory to describe the
striking fat-tailed nature of fluctuations throughout the Phanerozoic.

Despite the heterogeneity of explanations of Phanerozoic biodiversity,
consensus has emerged on one property of macroevolution: clades
experience different rates of morphological evolution, origination and
extinction \citep{simpson1953, sepkoski1984, gilinsky1994,
  rabosky2014}. Here we show that the simple fact of conserved rates
within clades and variable rates across clades is sufficient to
describe pervasive, fat-tailed fluctuations in taxonomic richness
throughout the marine Phanerozoic.  This biological mechanism has a
precise correspondence to the non-equilibrial theory from statistical
physics known as ``superstatistics'' \citep{beck2003} which has been
applied across the physical and social sciences \citep{beck2004,
  fuentes2009}. We leverage this correspondence to explain the
distribution of fluctuations in the standing richness of marine
invertebrates preserved in the Phanerozoic fossil record. We further
show that the specific mathematical form of this superstatistical
distribution is consistent with niche conservatism
\citep{roy2009range, hopkins2014} and pulsed exploration on an
adaptive landscape by higher taxa \citep{simpson1953,
  eldredgeGould1972, newman1985adaptive, hopkins2014}. We
operationally define ``adaptive landscape'' to mean a clade's set of
characteristics, and the fitness they impart to the clade, that
influences its macroevolution. Those characteristics could be
ecological (e.g.  substrate preference \citep{bambach1983, bush2007,
  hopkins2014}), morphological (e.g. body plan \citep{erwin2012}), or
macroecological (e.g. range size \citep{harnik2011,
  foote2008paleobiol}).

\subsection{Superstatistics of fossil biodiversity}

Superstatistics \citep{beck2003} proposes that non-equilibrial systems
can be decomposed into many local sub-systems, each of which attains a
unique dynamic equilibrium. The evolution of these dynamic equilibria
across sub-systems occurs more slowly. This separation in time scales
allows local systems to reach equilibrium while the system as a whole
is not \citep{beck2003}.  In the context of macroevolution we propose
that a clade with conserved macroevolutionary rates corresponds to a
sub-system in dynamic equilibrial.

In statistical mechanics, local sub-systems can be defined by a simple
statistical parameter $\beta$ often corresponding to inverse
temperature. In macroevolutionary ``mechanics'' we define the
$\beta_k$ of clade $k$ as the inverse variance of fluctuations $x_k$
in the number of genera within that clade, i.e. fluctuations in the
genus richness.  The $\beta_k$ thus represent the inverse variances,
what we term volatilities, of the origination-extinction processes of
genera with clades. The details of this origination-extinction
process, e.g. whether it is linear or subject to a carrying capacity,
are not of central importance to our analysis; so long as fluctuations
can reach a stationary distribution and are observed over
time-averaged intervals in a temporally coarse-grained fossil record
they will be approximately Gaussian (see Supplemental Section
\ref{sec:suppLimitDist}; \citep{grassmann1987}).

We make the hypothesis of dynamic equilibrium within a clade following
MacArthur and Wilson \citep{macWilson} in recognition that while the
identity and exact number of taxa will fluctuate stochastically from
random origination and extinction (taking the place of local
immigration and extinction in island biogeography), the overall
process determining the number of taxa, and by extension, fluctuations
in that number, is in equilibrium. Indeed, the different regions of
adaptive space occupied by different clades can be conceptualized as
islands with unique dynamic equilibria, albeit with macroevolutionary
processes determining the ``colonization'' of adaptive peaks, as
opposed to short timescale biogeographic processes.

The volatility of richness fluctuations will vary across these islands
in adaptive space as an emergent trait of a clade resulting from the
macroevolutionary fitness of the clade and the shape of the
surrounding adaptive landscape. Ultimately, volatility emerges from
the life histories, ecologies, and evolutionary histories that
drive each clade's macroevolutionary fitness and characterize its 
occupancy of different regions of an adaptive landscape. We do not
attempt to diagnose which characteristics of different regions account
for volatility differences, but others have found rates of origination
and extinction to depend on larval type \citep{jablonski2008}, body
plan \citep{erwin2012}, body size \citep{harnik2011}, range size
\citep{harnik2011, foote2008paleobiol}, and substrate preference
\citep{hopkins2014}. Not all of these traits would be considered
dimensions of an ecological niche or characteristics of a guild
\citep{bambach1983, bambach2007, bush2007}, but they all point to
different strategies that influence a clade's macroevolutionary
success. These characteristics result from interactions between
heritable traits and environments, which themselves may be
semi-heritable \citep{nicheCons}. Thus different regions of adaptive
space, and the clades occupying them, will experience different
magnitudes of stochastic fluctuations in taxonomic richness. As clades
occasionally split to fill new regions of adaptive space their pulsed
diversification determines the non-equilibrium nature of the entire
biota.

\subsection{Real paleontological data to test superstatistics}

To uncover the superstatistical nature of the marine invertebrate
Phanerozoic fauna we analyzed the distribution of fluctuations in
genus richness (the lowest reliably recorded taxonomic resolution)
using the Paleobiology Database (PBDB; {\tt paleobiodb.org}). We
corrected these raw data for incomplete sampling and bias using a new
approach described in the methods section. Occurrences from the PBDB
were matched to 49 standard time bins all of approximately 11MY
duration following previous publications \citep{alroy08,
  alroy2010}. Fluctuations in genus richness were calculated as the
simple difference between bias-corrected richnesses in adjacent
time bins.

To focus attention on the variance of fluctuations we zero-centered
each clade's fluctuation distribution. In this way we focus on
fluctuations about any possible trend toward net diversification or
extinction. Because ``equilibrium'' in the statistical mechanical
sense means a system undergoes coherent, concerted responses to
perturbation, the mean trend line (positive or negative) is of less
interest than deviations from it. We also note that the distributions
of fluctuations for most clades are already very close to a mean of 0
(mean at the family level: $0.038 \pm 0.176 \text{ SD}$), and so
centering has little influence on clade-specific fluctuation
distributions, consistent with the observation that origination is
often roughly equal to extinction \citep{foote2010Chapter}. Following
\citep{fuentes2009} we also ignore all instances of no change
(i.e. zero fluctuation).

We define potentially equilibrial sub-systems based on taxonomic
hierarchies as a full phylogenetic hypothesis for all marine
invertebrates is lacking.  Taxa ideally represent groups of organisms
that descend from a common ancestor and share similar ecologically and
evolutionary relevant traits \citep{mayr1965systZool, erwin2007}. Thus
our model assumes that at a given higher taxonomic level, within-taxon
fluctuations in richness are driven by equilibrial processes
characterized by Gaussian distributions. We further assume that new
higher taxa arise due to the emergence of sufficiently novel traits
(be they ecological, morphological, life history, or macroecological)
so that those new taxa occupy a new region of an adaptive
landscape. We lastly assume that different regions of adaptive space
are characterized by different volatilities in origination and
extinction.

To evaluate the optimal taxonomic level for sub-system designation, we
test our superstatistical theory using taxonomic levels from family to
phylum. Additionally, we compare our results to randomized taxonomies
and confirm that the observed fit of superstatistics is not an
artifact of arbitrary classification but instead represents real,
biologically relevant diversification processes within and between
clades. We find that families and orders conform to the assumptions of
our superstatistical model while classes and phyla do not.

\section{Results}

\begin{figure}[!h]
  \centering
  \includegraphics[scale=0.9]{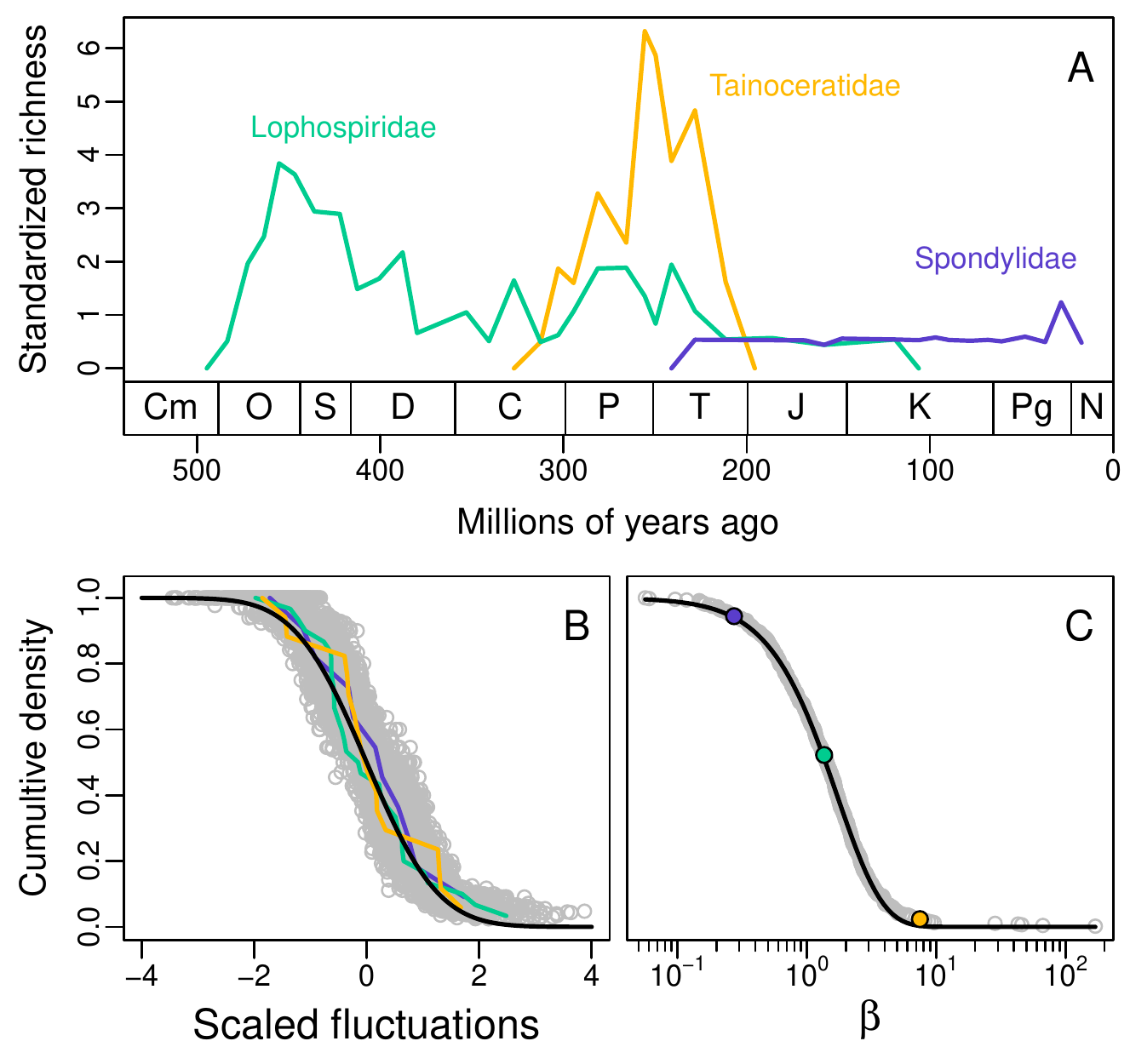}
  \caption[Variability in trajectories of within-family fluctuations
  in genus richness]{The distributions of within-family fluctuations
    in genus richness and across family volatilities. Panel (A) shows
    the richness fluctuation trajectories of three exemplar families
    and (B) shows an empirical cumulative density function of richness
    fluctuations aggregated across all families, highlighting the
    exemplar families. To display all families simultaneously we
    simply collapse their fluctuation distributions by dividing by
    their standard deviations. If families conform to the Gaussian
    hypothesis their scaled fluctuations should fall along the
    cumulative density line of a normal N(0, 1) distribution, as shown
    in (B). We further confirm this normal distribution in the
    supplement (Fig. \ref{figSupp:pkx_allTaxa}). In (C) the
    distribution of volatilities (inverse variances $\beta_k$) across
    all families matches very closely to a Gamma distribution (black
    line); exemplar families are again highlighted.}
  \label{fig:pk_f}
\end{figure}

We first evaluate the local equilibria of clades from family level to
phylum. We find that family level fluctuation distributions are well
approximated by Gaussians (Figs. \ref{fig:pk_f} and
\ref{figSupp:pkx_allTaxa}).  Three exemplar family-level dynamics are
highlighted in Figure \ref{fig:pk_f} to illustrate how different
volatility equilibria express themselves as actual richness
timeseries.  This Gaussian approximation also largely holds for
orders, but classes and phyla increasingly show deviations from
Gaussian with greater kurtosis corresponding to more frequent outliers
at these taxonomic levels (Fig. \ref{figSupp:pkx_allTaxa}).

To predict the superstatistical behavior of the entire marine
invertebrate Phanerozoic fauna we must integrate over all possible
local equilibria that each clade could experience. The stationary
distribution of $\beta_k$ values describes these possible equilibria,
specifying the probability that a given clade, chosen at random, will
occupy a region of adaptive space characterized by $\beta_k$.

We estimate the distribution of $\beta_k$'s simply as the maximum
likelihood distribution describing the set of volatilities for all
families, orders, classes, or phyla. Phanerozoic marine invertebrate
families clearly follow a Gamma distribution in their $\beta_k$ values
(Fig. \ref{fig:pk_f}). The Gamma distribution also holds for orders but
shows increasing deviations again for classes and especially phyla
(Fig. \ref{figSupp:fbeta_allTaxa}).

\begin{figure}[!h]
  \centering
  \includegraphics[scale=1]{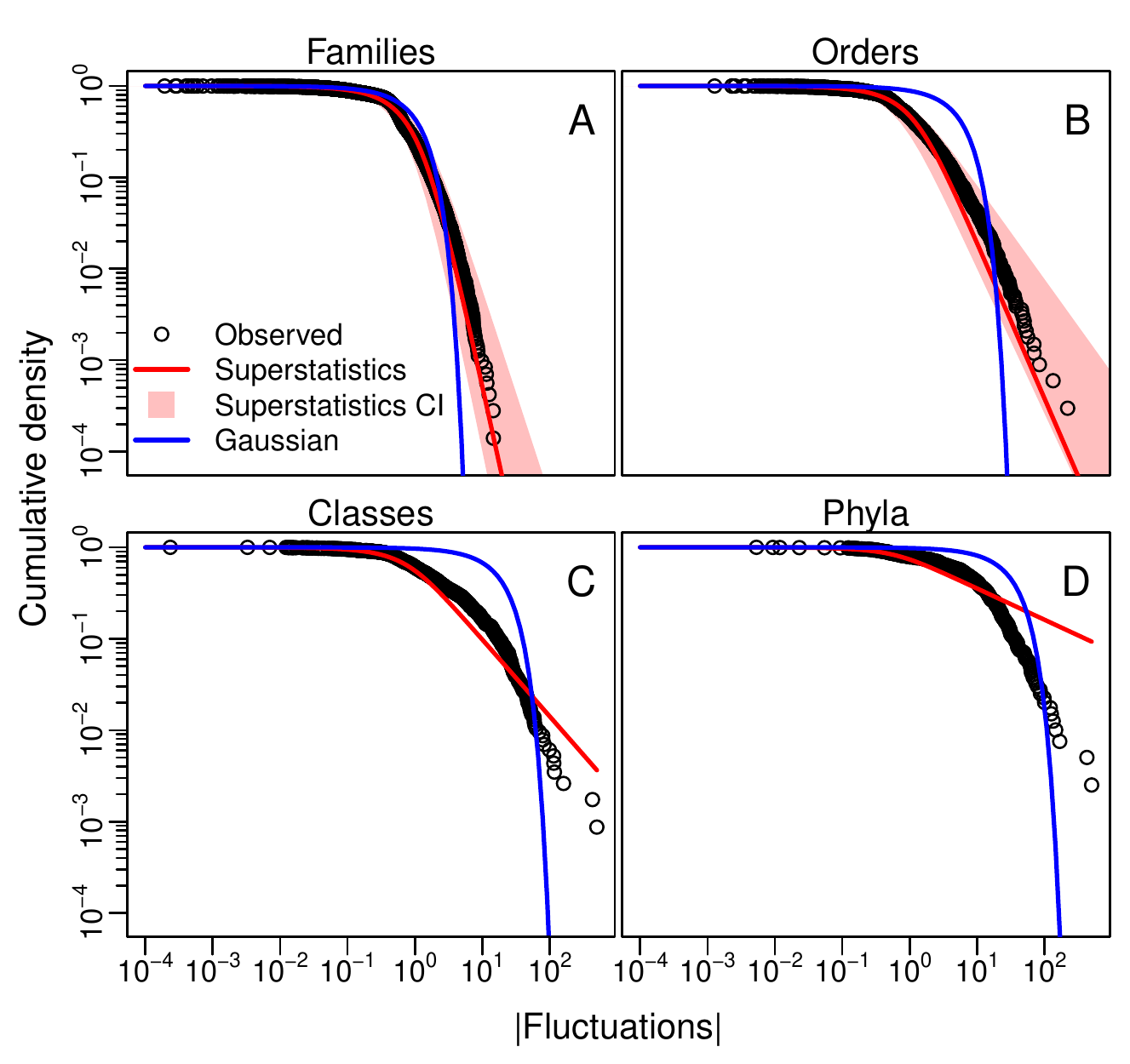} 
  \caption[Clade-level distribution of richness
  fluctuations]{Distribution of fluctuations in genus richness within
    different taxonomic groupings of marine invertebrates in the
    Paleobiology Database \citep{alroy08} after sampling
    correction. The distribution is fat-tailed as compared to the
    maximum likelihood estimate of the normal distribution (blue
    line).  At the family and order level the empirical distribution
    of fluctuations are well described by our superstatistical
    approach, both when computed from integrating over the
    distribution of observed variances (red line) and when fit via
    maximum likelihood (95\% confidence interval; red shading in
    (A) and (B)).}
  \label{fig:Px}
\end{figure}

Using the observation of within family statistical equilibrium and
Gamma-distributed $\beta_k$ parameters we can calculate, without
further adjusting free parameters, the distributions of family-level
fluctuations for the entire marine Phanerozoic, $P(x)$, as
\begin{equation}
  P(x) = \int_0^\infty p_k(x \mid \beta) f(\beta) d\beta \label{eq:PxInt}
\end{equation}
where
$p_k(x \mid \beta) = \sqrt{\frac{\beta}{2\pi}} e^{-\frac{\beta
    x^2}{2}}$ is the distribution of fluctuations within a family and
$f(\beta) = \frac{1}{\Gamma(b_1/2)}
\left(\frac{b_1}{2b_0}\right)^{b_1/2} \beta^{(b_1/2) - 1}
\text{exp}\left(-\frac{b_1 \beta}{2 b_0}\right)$ is the stationary
distribution of volatilities in richness fluctuations. The integral in
(\ref{eq:PxInt}) leads to
\begin{equation}
  \label{eq:Px}
  P(x) = \frac{\Gamma\left(\frac{b_1 +
        1}{2}\right)}{\Gamma\left(\frac{b_1}{2}\right)}
  \sqrt{\frac{b_0}{\pi b_1}} \left(1 + \frac{b_0
      x^2}{b_1}\right)^{-\frac{b_1 + 1}{2}}
\end{equation}
This corresponds to a non-Gaussian, fat-tailed prediction for $P(x)$
which closely matches aggregated family level fluctuations in the
bias-corrected PBDB (Fig. \ref{fig:Px}).

To quantitatively evaluate how well the superstatistical prediction
matches the family- and order-level data we constructed a 95\%
confidence envelope from bootstrapped maximum likelihood estimates of
$P(x)$. Observed fluctuations for both taxonomic levels fall within
these 95\% confidence envelopes (Fig. \ref{fig:Px}), indicating that
the data do not reject the superstatistical prediction. For further
comparison, we fit a Gaussian distribution to the observed
fluctuations, which corresponds to the equilibrium hypothesis that all
families conform to the same dynamic. Using Akaike Information
Criterion (AIC) we find that observed fluctuations are considerably
better explained by the superstatistical prediction than by the
Gaussian hypothesis ({\small $\Delta$}AIC = 1895.622). Thus, as
expected under the superstatistical hypothesis, the fat-tailed
distribution of fluctuations arise from the superposition of
independent Gaussian statistics of fluctuations within families.
Computing the distribution of aggregated fluctuations using orders
also closely matches the observed data (Fig. \ref{fig:Px}) but as we
further coarsen the taxonomy to classes and phyla we see increasingly
poorer correspondence between data and theory (Fig. \ref{fig:Px}).

\begin{figure}[!h]
  \centering
  \includegraphics[scale=1]{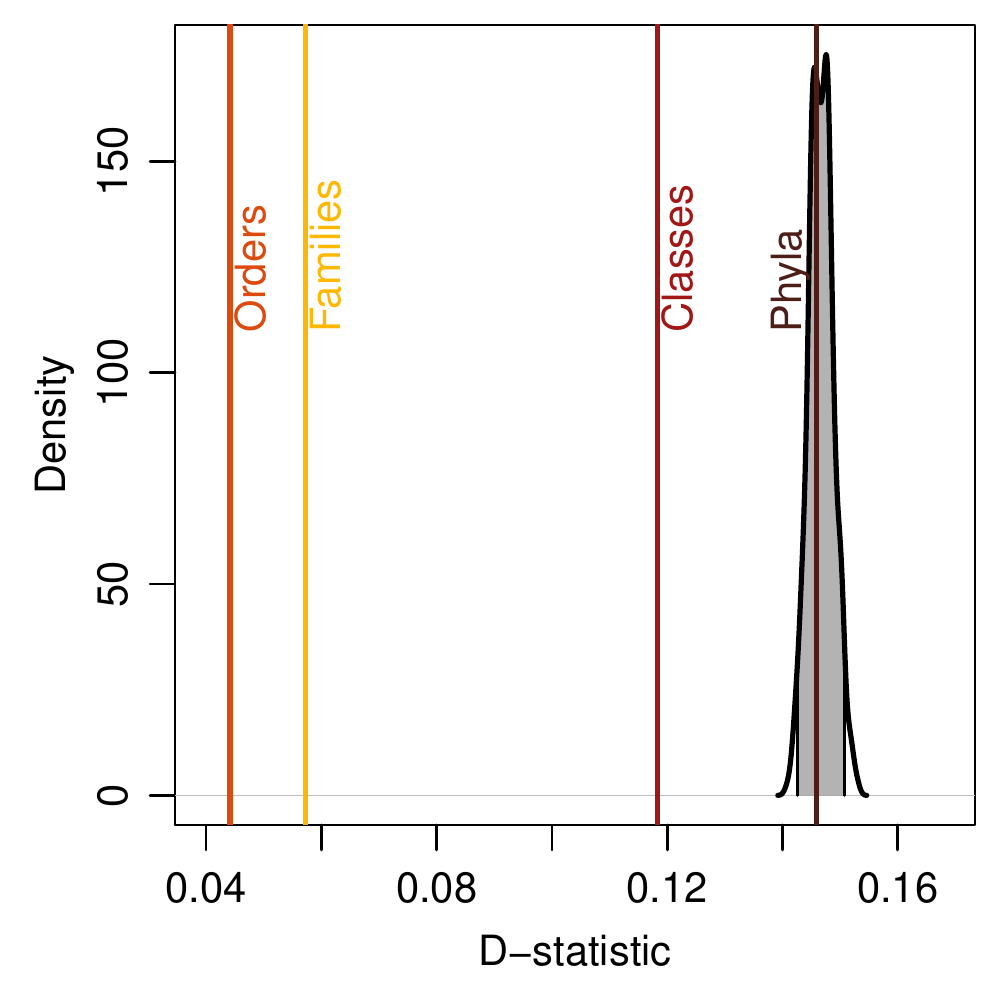}
  \caption[Goodness of superstatistical theory fit]{Distribution of
    Kolmogorov-Smirnov (KS) statistics from randomly permuting genera
    within families (gray shading represents 95\% confidence
    interval). Solid colored lines are observed KS statistics at
    different taxonomic levels as indicated.}
  \label{fig:dStat}
\end{figure}

We quantify this change in the goodness of fit with the
Kolmogorov-Smirnov statistic (Fig. \ref{fig:dStat}). We can see that
both families and orders have low Kolmogorov-Smirnov statistics, and
in fact order level designation of equilibrial subsystems performs
slightly better than the family level. Classes are substantially worse
and phyla worse yet with the Kolmogorov-Smirnov statistic of phyla
being no different than the null randomized taxonomies described
below.

However, if superstatistical theory explains the data, this worsening
fit with increasing taxonomic scale is expected as the different
classes and phyla should not represent dynamically equilibrial
sub-systems in their fluctuation dynamics. Instead, classes and phyla
aggregate increasingly disparate groups of organisms, and thus
effectively mix their associated Gaussian fluctuations, meaning that
one statistic should no longer be sufficient to describe class- and
phylum-level dynamics. We see this confirmed by the increasing
frequency of outlier fluctuations in within class and phylum level
fluctuation distributions (Fig. \ref{figSupp:pkx_allTaxa}). We can
also see that families and orders represent, on average, 1 to 2
ecospace hypercubes (defined by taxon environment, motility, life
habit, vision, diet, reproduction, and ontogeny \citep{bambach1983,
  bambach2007, bush2007}), respectively. In contrast, classes and
phyla represent, on average, 8 to 30 hypercubes, respectively
(Fig. \ref{figSupp:eeSpaceOcc}).

Our analysis indicates that orders and families are evolutionarily
coherent units with all subsumed taxa sharing key ecological
and evolutionary attributes allowing them to reach steady state
diversification independently from other clades at global scale. The
fact that both orders and families conform to theoretical predictions
is consistent with superstatistics. If superstatistics operates at the
order level, then the families subsumed by these orders should
represent random realizations of their order's stationary $\beta_k^{(order)}$
volatility. The sum of Gamma random variables is still Gamma, but with
new parameters, thus the family level distribution of
$\beta_k^{(family)}$ is still Gamma.

To further test the evolutionary coherence of families we conducted a
permutation experiment in which genera were randomly reassigned to
families while maintaining the number of genera in each family. For
each permutation, we calculated the superstatistical prediction and
its Kolmogorov-Smirnov statistic. The permutation simulates a null
model in which common evolutionary history is stripped away (genera
are placed in random families) but the total number of observed genera
per family is held constant. Because we ignore all instances of no
change (i.e. 0 fluctuation) we remove any possible large and
artificial gaps in the genus occurrences of these permuted
clades. Controlling for the total number of genera per family is key
because this could be purely an artifact of an arbitrary taxonomic
process \citep{yule1925, capocci2008} and genus richness alone could
be solely responsible for differences in the $\beta_k$ across clades.

We test the possibility that richness is responsible for variation in
$\beta_k$ in two ways. First, we find that the distribution of genus
richnesses within families is not itself distributed Gamma
(Fig. \ref{figSupp:richGamma}), indicating that there is not a simple
equivalence between $\beta_k$ and the richness of family $k$. Second,
we find that the number of genera in a family and that family's
$\beta_k$ value are indeed negatively correlated
(Fig. \ref{figSupp:betaByRich}). A negative correlation between clade
richness and $\beta_k$ is not unexpected because fluctuations are the
sums of the random variables representing genus origination and
extinction events; the more of these random variables in the summation
(i.e. the more genus richness in a clade) the higher the variance of
the summation.  Because $\beta_k \equiv 1/\sigma_k^2$ increasing
richness should lead to decreasing $\beta_k$ values. Thus we want to
know if this correlation accounts for all downstream superstatistical
results. The permutation test is specifically designed to determine if
the $\beta_k$ imposed by this correlation with richness are sufficient
to explain the observed superstatistical fit.

Repeating the null permutation of genera in families 500 times yields
a null distribution of Kolmogorov-Smirnov statistics that is far
separated from the observed values at the family and order levels
(Fig. \ref{fig:dStat}) suggesting that the good fit at these levels is
not merely a statistical artifact of classification or the richness of
clades, but carries important biological information. Classes approach
the null and phyla are no different. It should also be noted that the
width of 95\% confidence interval of this null distribution is not far
from the distance between the Kolmogorov-Smirnov statistics of orders
versus families, suggesting that differences of fit between these
taxonomic levels is at least partially accounted for by the randomness
of the sampling distribution of Kolmogorov-Smirnov statistics.

\section{Discussion}

%% why orders
Our analysis makes no assumption that orders and families should
correspond to superstatistical subsystems, but identifies them as the
appropriate level for marine invertebrates. Our study is the first to
demonstrate that complex patterns in the fluctuation of taxon richness
in the fossil record are the result of a simple underlying process
analogous to the statistical mechanisms by which complexity emerges in
large, non-equilibrium physical \citep{beck2004} and social systems
\citep{fuentes2009}.  We do so by identifying the biological scale at
which clades conform to locally independent dynamic equilibria in
fluctuations.  Equilibrium could result from many processes, including
neutrality \citep{macWilson, hubbell2001}, diversity-dependence
\citep{moen2014, foote2018} and processes that dampen---rather than
exacerbate---fluctuations in complex ecological networks
\citep{berlow2009}. These candidate processes are directly opposed to
the presumption of instability underlying the self-organized
criticality hypothesis for paleo biodiversity \citep{bak1993,
  sole1997}.

We show that the distribution describing the evolution to different
equilibria between orders and families is Gamma (Fig. \ref{fig:pk_f}).
A Gamma distribution, while consistent with multiple processes, could
result from evolution of diversification rates across an adaptive
landscape that promotes niche conservatism and pulsed exploration of
niche space \citep{cir1985}.  Specifically, if $\beta_k$ values are
associated with a clade's macroevolutionarily-relevant traits, and
those traits evolve via Ornstein-Uhlenbeck-like exploration of an
adaptive landscape, the resulting stationary distribution of $\beta_k$
will be Gamma \citep{cir1985}. For macroevolutionary rates to vary in
a way consistent with the observed superstatistical description of
fluctuations this landscape cannot be flat (i.e. equal fitness
everywhere), but instead must be rugged. Thus, niche conservatism
around local fitness optima in adaptive space interrupted by adaptive
exploration is likely \citep{newman1985adaptive,
  gavrilets2004book}. The specifics of how this adaptive landscape is
shaped and is traversed by evolving clades determine the exact form of
the distribution of $\beta_k$ volatilities, in the case of the marine
Phanerozoic resulting in a Gamma distribution. Our work thus motivates
further study of the trait spaces and evolutionary shifts consistent
with Gamma-distributed equilibria in richness fluctuation
volatilities.

We show that the pulsed shift to different equilibria between orders
and the families they subsume is sufficient to explain the
characteristically fat-tailed distribution of richness fluctuations
when the marine Phanerozoic invertebrate fauna is viewed as a whole
macrosystem.  Armed with an understanding of the statistical origin of
this diversification pattern we can explore which models of niche
conservatism and pulsed adaptive radiation are consistent with the
statistical behavior of the Phanerozoic. Our statistical theory
provides new motivation for identifying the eco-evolutionary causes of
innovations between lineages and how those innovations are eventually
conserved within lineages. Using the superstatistical prediction as a
theoretical baseline, we can also go on to identify and robustly
examine the mechanisms underlying deviations from statistical
theory. For example, some clades wax and wane systematically, and
possibly non-symmetrically, through time \citep{liow2007,
  foote2008paleobiol, quental2013}, a pattern that we cannot explain
with superstatistics alone.

Superstatistics could also be applied to other areas of evolution and
macroecology.  For example new phylogenetic models already consider
heterogeneous rates of diversification (e.g., \citep{rabosky2014}) as
expected between different subsystems. The superstatistics of clades
in adaptive landscapes could motivate models that jointly predict
changes in traits and diversification, a research area currently
struggling with model inadequacy \citep{rabosky2017fisse}. This
framework could also provide a new paradigm in modeling the
distributions of richness, abundance, and resource use in non-neutral
communities which can be viewed as emerging from the combination of
locally equilibrium subsystems. Non-neutral models in ecology are
criticized for their over-parameterization \citep{rosindell2011}, yet
a persistent counter argument to neutral theory \citep{hubbell2001} is
the unrealistic assumption of ecological equivalency and poor
prediction of real dynamics \citep{rosindell2011}. If ecosystems are
viewed as the superposition of many individualistically evolving
clades, each exploiting the environment differently and thus obeying a
different set of statistics, then diversity dynamics could be
parsimoniously predicted with superstatistics while incorporating real
biological information on ecological differences between taxa.

Superstatistics is a powerful tool to derive macro-scale predictions
from locally fluctuating sub-systems whose evolution is driven by
interesting, but complex and difficult to model, biological
mechanisms. As such, applications of superstatistics to a wide variety
of patterns in ecological and evolutionary systems are ripe for
exploration.

\section{Methods and Materials}

All data processing and analyses were preformed in R \citep{rcite} and
all code needed to reproduce our study are provided, with added
explanation, in supplemental Appendix A.

\subsection{Paleobiology Database data download and filtering}
Data on individual fossil occurrences and the ecospace characteristics
of Phanerozoic marine invertebrates were downloaded from the
Paleobiology Database (PBDB; \newline\texttt{https://paleobiodb.org}) on 16
November 2018 via the database's API (data retrieval and processing
script available in the supplement). Collections were filtered using
the same approach as Alroy \citep{alroy08} to insure that only well
preserved marine invertebrate occurrences were used in subsequent
analyses. This filtering resulted in 815,222 unique genus-level
occurrences. These were further filtered to exclude those occurrences
without family-level taxonomy and those collections with age estimate
resolutions outside the 11MY time bins proposed by Alroy
\citep{alroy08} resulting in 454,033 occurrences. Time bins were
compiled from {\tt http://fossilworks.org} with a custom script
reproduced in the supplement. The first and last of these time bins,
corresponding to the earliest Cambrian and the latest Cenozoic, were
excluded from analysis because their sampling completeness (see below)
could not be assessed.

\subsection{Correcting for imperfect and potentially biased sampling}
\label{sec:3TP}
We use a new and flexible method to correct for known sampling
incompleteness and biases in publication-based specimen databases
\citep{alroy08, alroy2010}. Incompleteness is inherent in all
biodiversity samples, the fossil record being no exception
\citep{miller1996, foote2016, starrfelt2016, close2018}.  In addition
to incompleteness, bias may result from preferential publication of
novel taxa \citep{alroy2010} which exacerbates the difference between
poorly-sampled and well-sampled time periods. We therefore develop a
simple two-step method: we first correct genus richness for incomplete
sampling using the ``three-timer'' correction \citep{alroy08} and then
further correct this three-timer richness estimate by accounting for
any correlation between the number of genera and the number of
publications in a time period.

The three-timer correction estimates the probability of failure to
observe a genus in a given time period $p_t$ as the number of times
any genus is recorded before and after that period but not during,
divided by the number of genera whose occurrence histories span the
period $t$.  To calculate the sampling-corrected richness
$\hat{D}_{kt}$ of a clade $k$ in the time period in question, the
observed genera within that clade and time period are divided by
$1 - p_t$ and their occurrences summed:
\begin{equation}
  \hat{D}_{kt} = \sum_{j \in k} \frac{I_{jt}}{1 - p_t}
\end{equation}
where $j \in k$ designates genera in clade $k$ and $I_{jt}$ is an
indicator equal to 1 if genus $j$ occurs in time period $t$.

$\hat{D}_{kt}$ is the maximum likelihood estimator of richness in a
simple occupancy through time type model assuming binomial sampling
\citep{royleDorazio}, and in that way mimics other proposed methods
for the fossil record \citep{foote2016, starrfelt2016}. We avoid
parametrically modeling the sampling process through time by instead
taking a sliding window of time bins from the Cambrian to the
Cenozoic. It should be noted that the three-timer correction compares
favorably to other similar methods to account for imperfect detection
\citep{alroy2014}

To eliminate further bias due to preferential publication of novel
taxa \citep{alroy2010} we divide the three-timer-corrected number of
genera per family per time period by the expected number of genera
given publications in that time period.  The expected number is
calculated by regressing the log-transformed three-timer-corrected
number of genera on log-transformed number of publications. There is
only a weak trend toward higher richness with more publications
(Fig. \ref{figSupp:divByPub}) meaning that the most important correction
comes from the three timer correction.

Our new method re-scales each genus occurrence from 0 or 1 (absent or
present) to a weighted number continuously ranging between 0 and
1. Because these weighted numbers represent sampling and
bias-corrected {\it occurrences} we can add them arbitrarily,
corresponding to the membership of any given genus in any given higher
taxonomic group.  We must, however, choose a taxonomic level at which
to evaluate the relationship between richness and publications; we
choose the level of family because this is the most finely resolved
option.

We opt not to use subsampling methods \citep{miller1996, alroy2010,
  kocsis2018} because these approaches would not be advisable for
clades with few genera. However, our new method achieves similar
results to subsampling procedures at the global scale across all
clades. We directly compare our predicted time series of global
fluctuations in genus richness with results derived from rarefaction
and shareholder quorum subsampling (SQS) in Figure
\ref{figSupp:3TPub}.  Our method shows very minor differences with
these subsampling-based predictions and any discrepancies do not
impact the statistical distribution of fluctuations
(Fig. \ref{figSupp:3TPub}).

\subsection{Superstatistical methods} \label{sec:numMeth}

We first derive the superstatistical distribution $P(x)$ by fitting
Gaussian distributions to clade-level distributions of fluctuations
$p_k(x)$, extracting the inverse variances $\beta_k$ of those
$p_k(x)$, testing the best function to describe the distribution of
$\beta_k$, and then integrating
$P(x) = \int_{\beta}p_k(x | \beta) f(\beta)$. This process allows no
free parameters to hone the fit of $P(x)$ to the data.  However, each
inverse variance must of course be estimated for each clade, making
its good fit to data all the more surprising.  To do so we use least
squares instead of maximum likelihood because the asymmetric
fluctuation distributions of small clades were more reliably fit with
curve fitting than with maximum likelihood.

We also estimated $P(x)$ directly from the family-level data using
maximum likelihood to compare the fit of our superstatistical
prediction and that of a simple Gaussian distribution using AIC. To
calculate a likelihood-based confidence interval on our prediction we
bootstrapped the data, subsampling fluctuations with replacement from
all families and fit superstatistics using maximum likelihood to the
aggregated fluctuation distribution of each bootstrap replicate.

\bibliographystyle{science-advances}

\section*{Acknowledgments}
\begin{itemize}
\item[{\bf General:}] We thank John Harte, Rosemary Gillespie, Linden
  Schneider, Jun Ying Lim, and David Jablonski for helpful
  discussion. We thank Aaron Clauset and four anonymous reviewers for
  greatly improving the quality of this manuscript. We thank the many
  contributors to the Paleobiology Database for making data available.
\item[{\bf Funding:}] AJR thanks funding from the Fulbright Program
  (USDoS Bureau of Educational and Cultural Affairs), the National
  Science Foundation Graduate Research Fellowship Program and the
  Omidyar Program at the Santa Fe Institute; MAF thanks FONDECYT
  1140278; PM thanks CONICYT AFB170008, ICM-P05-002 and FONDECYT
  1161023.
\item[{\bf Author contributions:}] AJR, MAF and PAM designed the
  study; AJR and MAF preformed the analyses; AJR, MAF and PAM
  interpreted the results and wrote the manuscript.
\item[{\bf Competing interests:}] none.
\item[{\bf Data and materials availability:}] Data are available
  through the Paleobiology Database ({\tt paleobiodb.org}) and all
  code needed to interface with the {\tt paleobiodb.org} API, process,
  clean, and ultimately analyze the data are available online at \\
  {\tt github.com/ajrominger/paleo\_supStat}. This github repository
  also hosts the exact download from {\tt paleobiodb.org} used in this
  analysis. All required scripts are also available and explained in
  supplemental Appendix A.
\end{itemize}

\clearpage

\newcommand{\beginsupplement}{%
  \setcounter{table}{0}
  \renewcommand{\thetable}{S\arabic{table}}%
  \setcounter{figure}{0}
  \renewcommand{\thefigure}{S\arabic{figure}}%
  \setcounter{section}{0}
  \renewcommand{\thesection}{S\arabic{section}}%
}

\beginsupplement

\begin{center}
{\LARGE \bf Supplementary materials}
\end{center}
\vspace{2em}

\section{Limit distribution of a time-averaged homogeneous
  origination-extinction process}
\label{sec:suppLimitDist}

Fossil taxa gain and lose genera according to an
origination-extinction process. We, however, do not see every event of
this processes but rather a time average imposed by the coarse
resolution of the rock record. In our analysis we use time bins of
approximately 11 MY and it is over this duration that the history of
originations and extinctions are time-averaged resulting in observed
taxon richnesses and fluctuations thereof. Such time-averaged Markov
processes have already been shown to be asymptotically
Gaussian\citep{grassmann1987}.  Using the asymptotic Gaussian
approximation is also a more appropriate distribution for our sampling
and bias-corrected richness estimates because these estimates are not
integer-valued but rather continuous random variables.

What is more, because preservation and sampling are far from complete
we likely only recover taxa when they are in an abundant and largely
stationary period in their macroevolution \citep{liow2007}. This
stationarity gives us another lens on the asymptotic normality of
fluctuations because average per capita rates of origination and
extinction would be equal (i.e.  $\lambda = \mu \equiv \rho$) over a
coarse-grained interval of duration $\tau$ and the number of
origination or extinctions events (call such events $Y$) each follow
an inhomogeneous Poisson process with rate $\tau \rho N_t$. Here $N_t$
is the time-averaged number of genera in the taxon of interest during
the interval of length $\tau$ at time $t$.

The difference of these Poisson distributions is again asymptotically
Gaussian.  Our analysis does not depend on all clades being perfectly
stationary with $\lambda = \mu$ because of the asymptotics of
time-averaged Markov processes. Indeed we zero-center all fluctuation
time series to avoid possible net diversification or extinction from
biasing our analysis of fluctuation volatilities.

\section{Evaluation of sampling bias correction methods}
\label{sec:suppBiasEval}

\begin{figure}[!hp]
  \centering
  \includegraphics[width=0.4\textwidth]{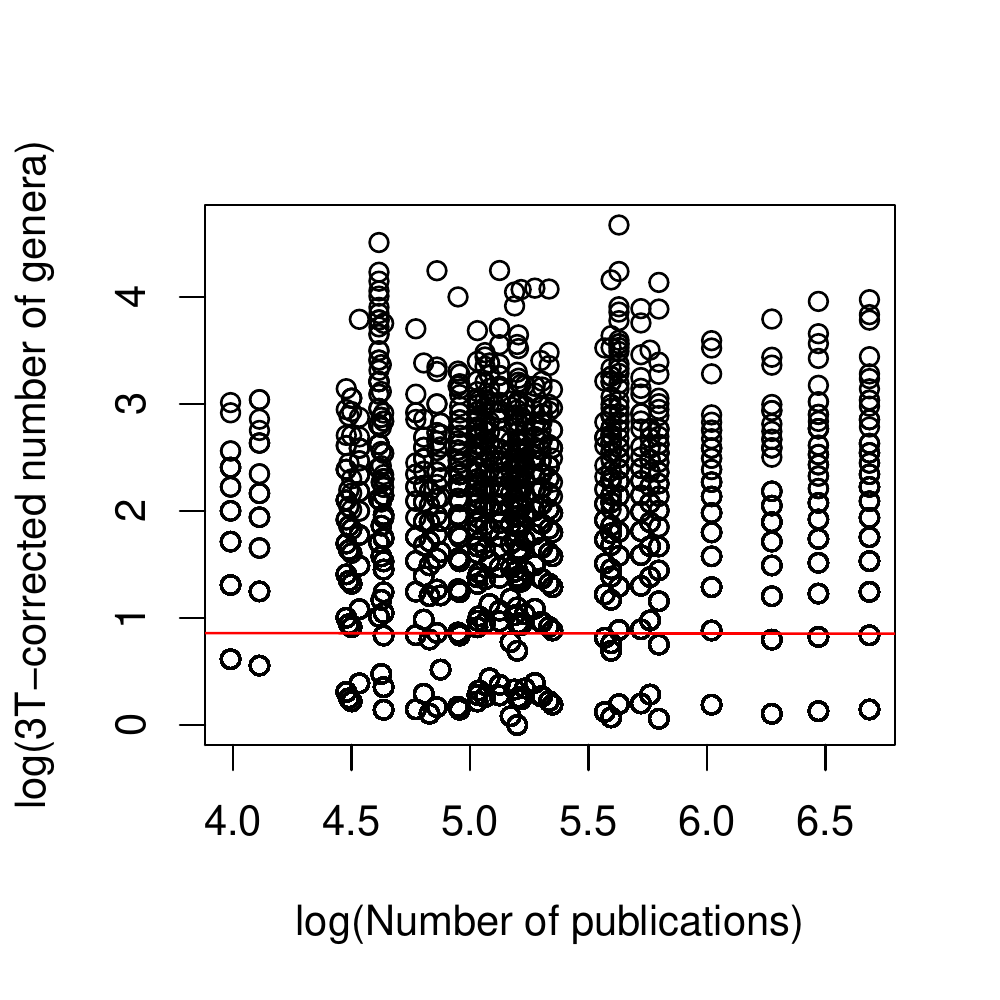}
  \caption{Relationship between number of publications and three-timer
    (3T) corrected genus richness at the family level as recorded by
    the PBDB.}
  \label{figSupp:divByPub}
\end{figure}

Our sampling and bias-correction method first accounts for imperfect
detection within a binomial sampling framework as described in the
main text, and then further corrects for potential publication bias
using simple log-log regression.  We reproduce that regression of
log-richness versus log-number of publications here
(Fig. \ref{figSupp:divByPub}). 

\begin{figure}[!hp]
  \centering
  \includegraphics[width=0.7\textwidth]{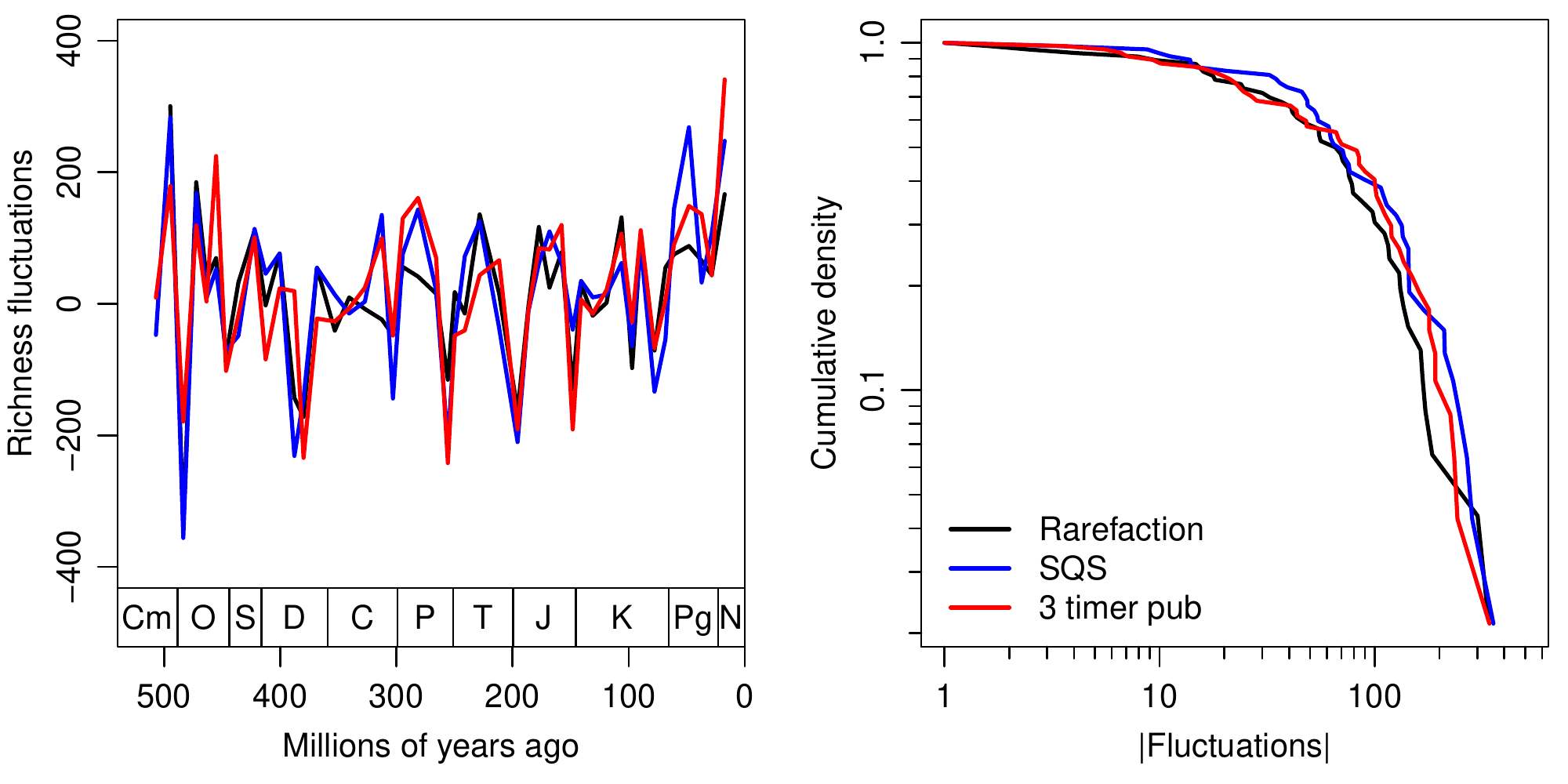}
  \caption{Comparison of rarefaction (black line) and SQS (blue)
    with our three-timer and publication bias correction method
    (red). The time-series of all marine invertebrate genera shows
    general agreement with the only major deviations toward the modern
    (A). Despite these differences the distribution of fluctuations in
    genus richness across all marine invertebrates show good agreement
    (B).}
  \label{figSupp:3TPub}
\end{figure}

We compare our sampling and bias-correction method to other more
established approaches. Specifically we use the newly available R
package {\it divDyn} \citep{kocsis2018} to produce subsampling-based
richness estimates for the Phanerozoic timeseries of marine
invertebrates. In Figure \ref{figSupp:3TPub} we compare classical
rarefaction and shareholder quorum subsampling (SQS) with our
method. All samples were rarified to 120 occurrences, which is
approximately the maximum possible rarefied sample size across all
time bins, and the SQS quorum was set to 0.75 to similarly approximate
this common sampling denominator across time bins. For both
rarefaction and SQS we averaged 50 subsampled replicates.

\section{Understanding deviations from superstatistics at higher
  taxonomic levels}
\label{sec:suppSstatTaxLevels}

\begin{figure}[!hp]
  \centering
  \includegraphics[width=0.8\textwidth]{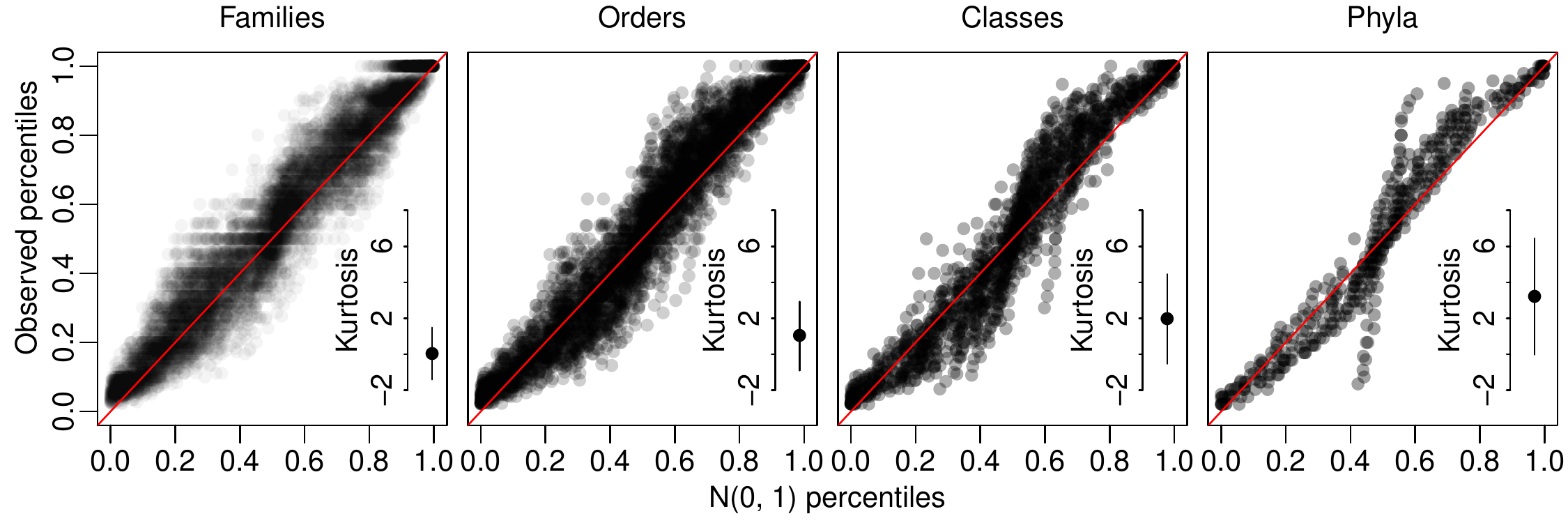}
  \caption{Change in within clade richness fluctuation distributions
    with increasing taxonomic level. The percentile-percentile plots
    show how the percentiles of observed re-scaled fluctuation
    distributions compare to expected percentiles from a Gaussian
    distribution with mean 0 and variance 1. We can see that families
    and order conform to a linear relationship, albeit with the later
    showing some signs of an s-shaped pattern. Clases and phyla show
    stronger deviations from the linear trend with a marked s-shaped
    relationships. Inset plots show how kurtosis increases from 0 (the
    value for a Gaussian distribution) at the family level to
    increasingly larger values at higher taxonomic levels.}
  \label{figSupp:pkx_allTaxa}
\end{figure}

\begin{figure}[!hp]
  \centering
  \includegraphics[width=0.8\textwidth]{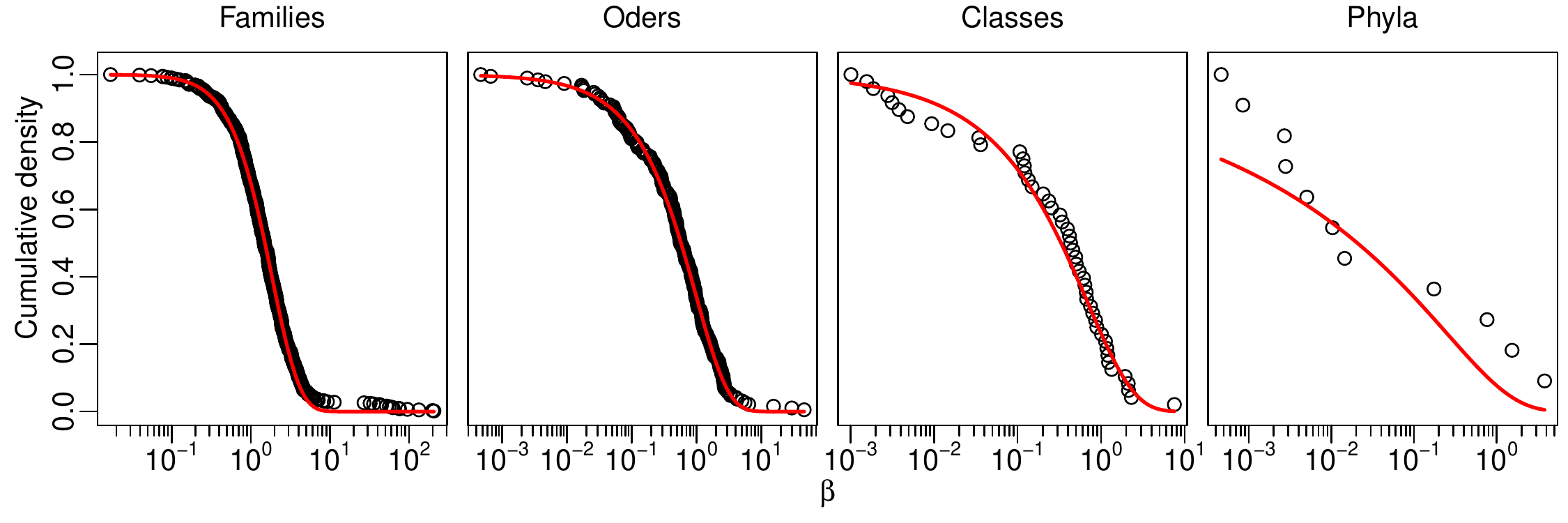}
  \caption{Change in the distributions of $\beta_k$ across clades of
    increasing taxonomic level. Points are observed $\beta_k$ values
    and red lines are the best-fit Gamma distributions. Deviations
    increase particularly at the class and phylum levels.}
  \label{figSupp:fbeta_allTaxa}
\end{figure}

To explore why deviations from super statistics increase with
increasing taxonomic level we explore how the distributions of
richness fluctuations $p_k(x | \beta_k)$ and fluctuation volatilities
$f(\beta_k)$ change with changing taxonomic level. We find that
richness fluctuation distributions experience increasing frequencies
of outliers (increasing kurtosis) with higher taxonomic level
(Fig. \ref{figSupp:pkx_allTaxa}). We also find that observed
fluctuation volatility distributions increasingly depart from a Gamma
distribution at the levels of classes and phyla
(Fig. \ref{figSupp:fbeta_allTaxa}).

\section{Ecospace occupation of higher taxa}
\label{sec:suppGuilds} 

\begin{figure}[!hp]
  \centering
  \includegraphics[width=0.4\textwidth]{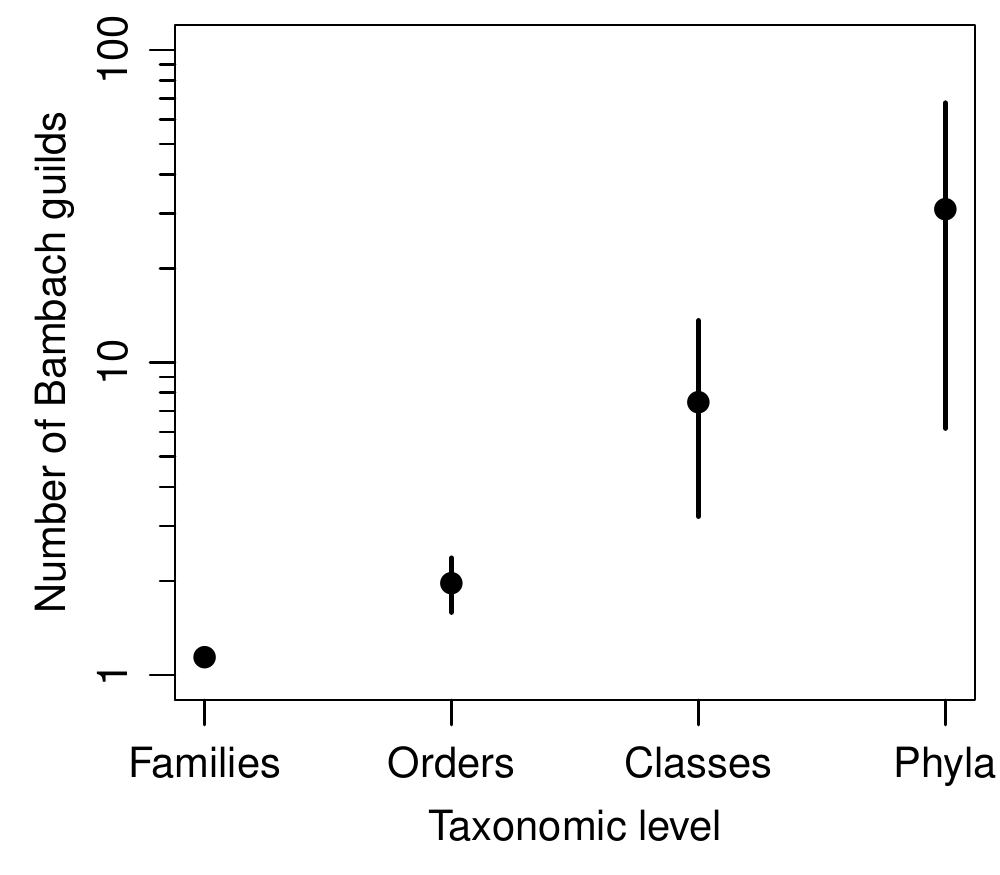} 
  \caption{Relationship between number of ecospace hypercubes occupied
    and taxonomic level.}
  \label{figSupp:eeSpaceOcc}
\end{figure}

We posit that part of the increasing divergence between
superstatistics and observed fluctuations and the increase in
fluctuation outliers at higher taxonomic levels is that these higher
taxa increasingly aggregate disparate types of organisms. One way to
evaluate this idea is to count the ecospace hypercubes
\citep{bambach1983, bambach2007, bush2007} occupied by taxa at
different levels. We use the ecological characteristics reported by
the PBDB: taxon environment, motility, life habit, vision, diet,
reproduction, and ontogeny. In Figure \ref{figSupp:eeSpaceOcc} we find
that families comprise, on average, 1 hypercube, orders comprise 2
hypercubes on average, and classes and phyla comprise many more.

\section{Relationship between $\beta_k$ and clade richness}
\label{sec:suppBetaRichness} 

\begin{figure}[!hp]
  \centering
  \includegraphics[width=0.4\textwidth]{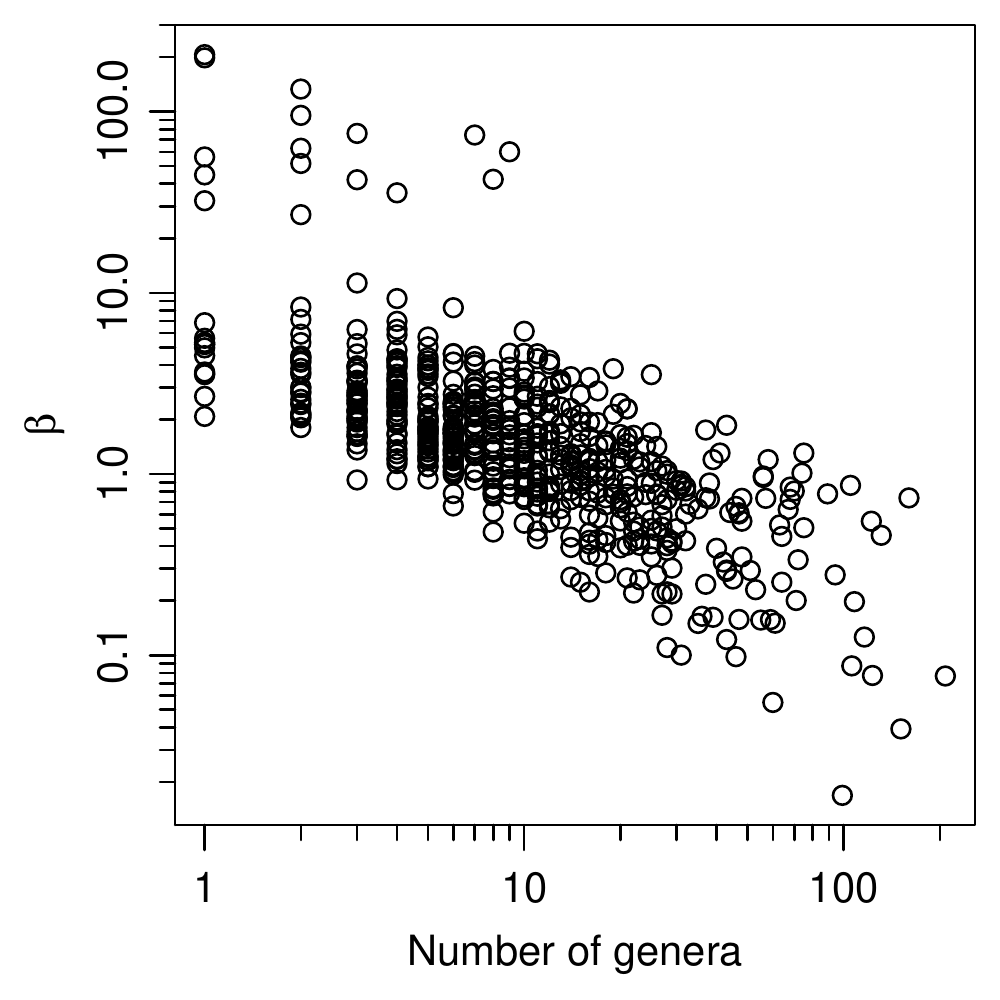} 
  \caption{Relationship between fluctuation volatility $\beta_k$ and
    genus richness at the family level.}
  \label{figSupp:betaByRich}
\end{figure}

There is likely to be a relationship between richness of clade $k$ and
its fluctuation volatility $\beta_k$ because both extinction and
origination (i.e. the formation of new genera) contribute to
volatility. Thus we expect that higher variance in richness
fluctuations (i.e. smaller $\beta_k = 1/\text{variance}$) will be
correlated with higher richness.  Indeed, Figure
\ref{figSupp:betaByRich} shows this to be true. 

\begin{figure}[!hp]
  \centering
  \includegraphics[width=0.4\textwidth]{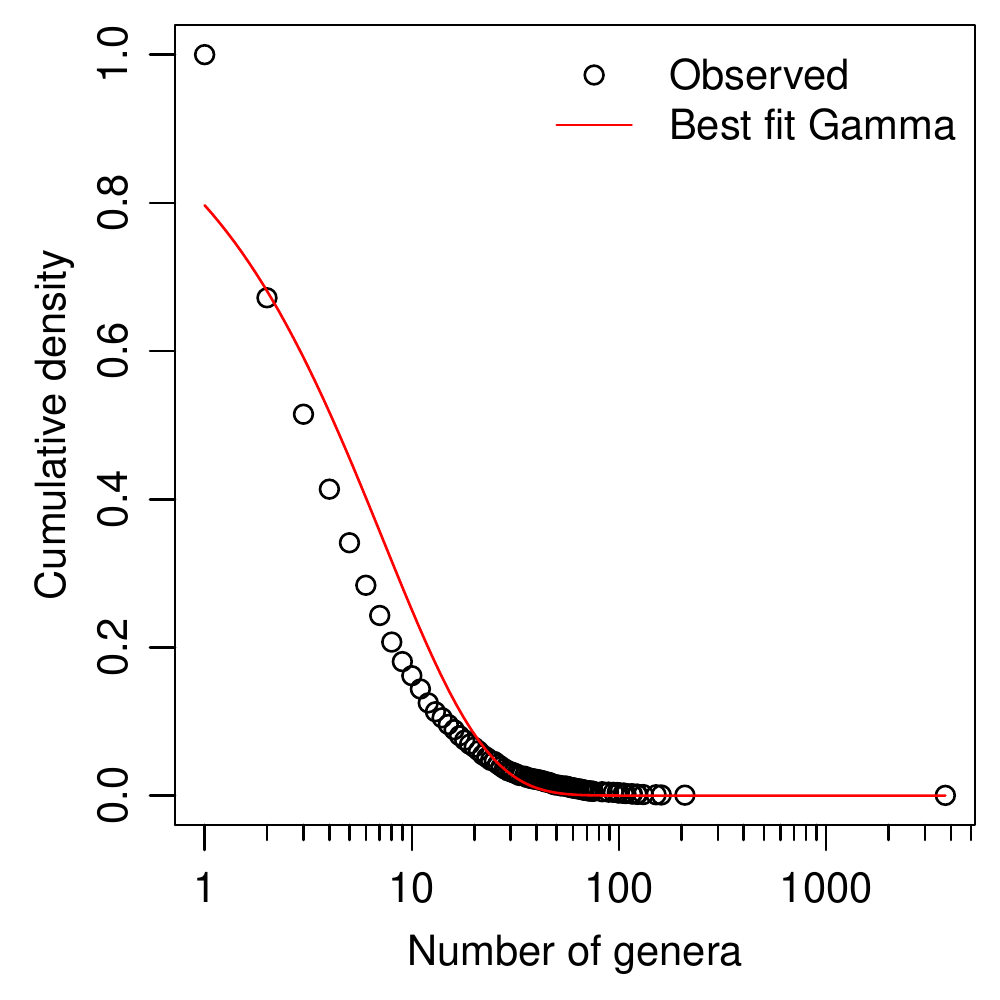} 
  \caption{Distribution of genus richnesses within families. Red line
    shows best-fit Gamma distribution which clearly deviates from the
    observed cumulative density function (black points).}
  \label{figSupp:richGamma}
\end{figure}

In the main text we use permutation to evaluate whether this
correlation is responsible for the observed good fit of
superstatistics, and find that this correlation alone is not
sufficient. In addition to this permutation test, we directly evaluate
whether the distribution of clade richness at the family level
conforms to a Gamma distribution (Fig. \ref{figSupp:richGamma}). If
the family-level richness distribution had mirrored the distribution
of $\beta_k$ values this may have suggested that richness was largely
responsible for superstatistical behavior. However, we find that
family richness is not Gamma (Fig. \ref{figSupp:richGamma}),
reaffirming the permutation-based findings that the $\beta_k$ values
derive from more nuanced biological mechanisms.

\end{document}